\documentclass[10pt, english,prl,aps,twocolumn,floatfix,superscriptaddress]{revtex4-1}
\usepackage[T1]{fontenc}
\usepackage{color}
\usepackage{babel}
\usepackage{verbatim}
\usepackage{amsthm}
\usepackage{amsmath}
\usepackage{graphicx}
\usepackage{color}
\def\ket#1{{|#1\rangle}}

\usepackage[pdfstartview={FitH},bookmarks=false]
 {hyperref}

\theoremstyle{plain}

  \theoremstyle{remark}

  \providecommand{\remarkname}{Remark}
\providecommand{\theoremname}{Theorem}

\newcommand{\Uexp}{U_{e}}
\newcommand{\eps}{\epsilon}
\newcommand{\pol}{P}

\makeatletter

\begin{document}

\title{Experimental Saturation of the Heat-Bath Algorithmic Cooling bound}

\author{Sebastian Zaiser}
\affiliation{3. Physikalisches Institut, Center for Applied Quantum Technologies and IQST, University of Stuttgart, Stuttgart, Germany}
\author{Billy Masatth}
\affiliation{Department of Physics, Chinese University of HongKong, Shatin, HongKong, China}
\author{D. D. Bhaktavatsala Rao}
\email{d.dasari@physik.uni-stuttgart.de}
\affiliation{3. Physikalisches Institut, Center for Applied Quantum Technologies and IQST, University of Stuttgart, Stuttgart, Germany}
\affiliation{Max Planck Institute for Solid State Research, Stuttgart, Germany}
\author{Sadegh Raeisi}
\email{sadegh.raeisi@gmail.com}
\affiliation{Department of Physics, Sharif University of Technology, Tehran, Iran}
\author{J\"{o}rg Wrachtrup}
\affiliation{3. Physikalisches Institut, Center for Applied Quantum Technologies and IQST, University of Stuttgart, Stuttgart, Germany}
\affiliation{Max Planck Institute for Solid State Research, Stuttgart, Germany}

\begin{abstract}
Heat-Bath Algorithmic cooling (HBAC) techniques provide ways to
selectively enhance the polarization of target quantum subsystems. 
However, the cooling in these techniques are bounded. Here we report the first 
experimental observation of the HBAC cooling bound. 
We use HBAC to hyperpolarize nuclear spins in diamond. 
Using two carbon nuclear spins as the source of polarization (reset) and the ${}^{14}N$ 
nuclear spin as the computation bit, we demonstrate that 
repeating a single cooling step increases the polarization 
beyond the initial reset polarization and reaches the cooling 
limit of HBAC. We benchmark the performance of our experiment over a range of 
variable reset polarization.  
With the ability to polarize the reset spins to different 
initial polarizations, we envisage that the proposed model could serve as a test bed for studies on Quantum Thermodynamics.
\end{abstract}
\maketitle

There is a rapid growth in the scope and sophistication of spin-based quantum technologies
ranging from quantum-information applications to quantum sensing\cite{paolarev}. 
In a large number of cases the quantum spin system represent a central spin system, i.e. a central well controllable spin, which in most cases is an electron spin that dominates the dynamics of its environment. Irrespective of the particular application, controlling this environment is key for proper function of the quantum system \cite{qec}. Essentially one can try to eliminate all spins from the environment to enhance the central spin coherence time by dedicated material growth, e.g. isotope purification \cite{isotope}. Alternatively the spin environment can be polarized as to not decohere the central spin \cite{DNPNV}.

\begin{figure}
\begin{centering}
\includegraphics[width=1.0\columnwidth]{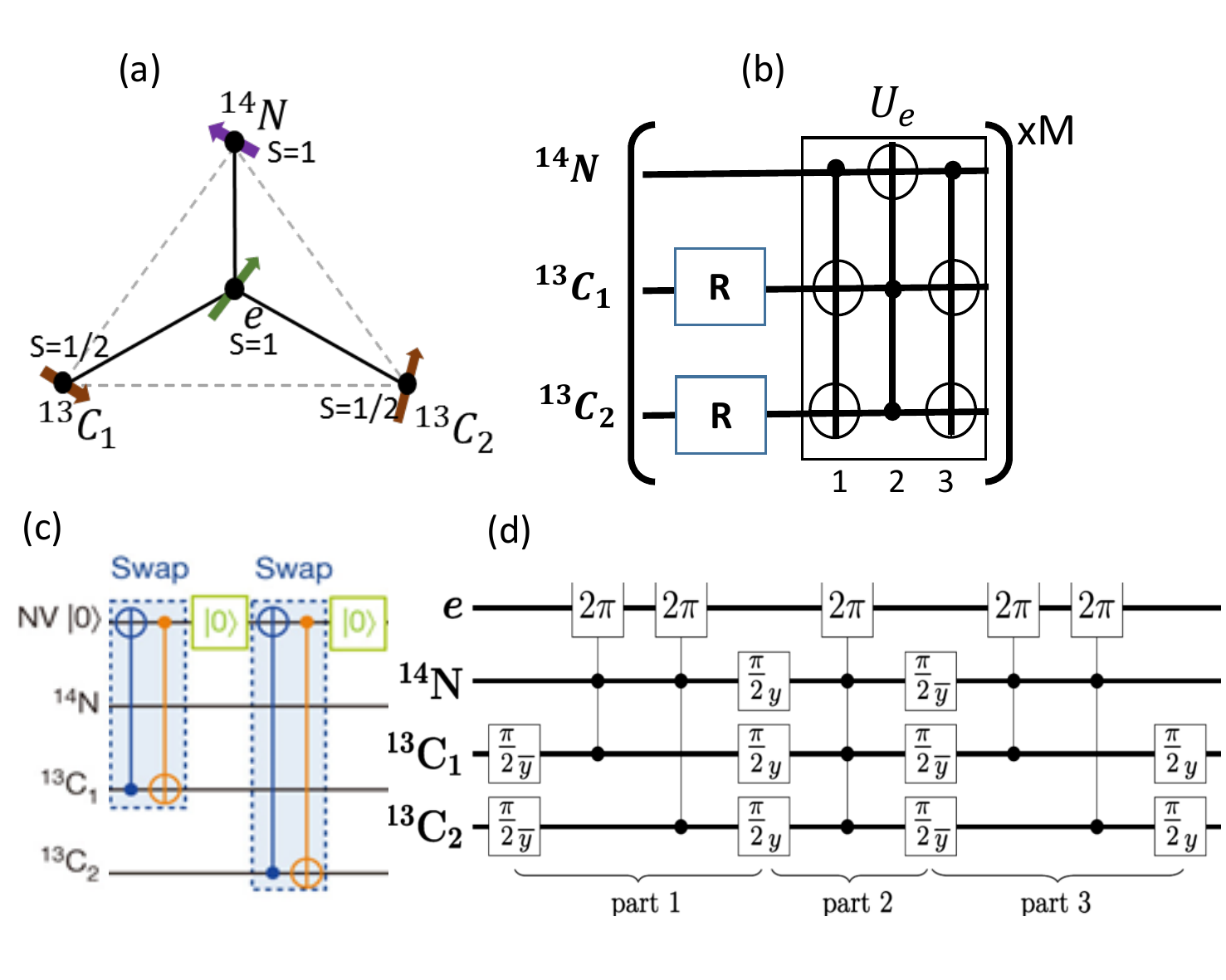} 
\par\end{centering}
\caption{\label{fig:Scheme}(a) Schematic representation of our physical system consisting of three nuclear spins $(14N, 13C, 13C)$ and a central electron spin (e) to which they are coupled. 
Various quantum gates among the n-spins are mediated by the electron. (b) The quantum circuit implemented for HBAC. 
The power indicates that this is an iteration and is repeated $M$ times. (c)-(d) Experimental Quantum circuits for implementing the reset operation, R and $U_e$ in (b).
 }
\end{figure}

Over the past decade single nitrogen-vacancy (NV) centers in diamond have emerged as a novel atomic-size magnetometer for detecting nuclear spin ensembles or even single nuclear spins with high sensitivity \cite{amitrev}. 
Essentially it is a prototype central spin system, consisting of a substitutional nitrogen atom and an adjacent vacancy. 
Its spin triplet ($S=1$) ground state can be polarized and read out optically, so that electron spin resonance experiments can be performed on spin ensembles or single spins \cite{nvrev}. The electron spin associated with the NV center is a very good qubit and at the same time allows control over its immediate nuclear spin environment, e.g.  the associated ${}^{14}N$ or the 13C nuclear spins  in the proximity of the NV center \cite{qec}. In this work we use these nuclear spins, and a combined usage of the HH and HBAC cooling schemes to demonstrate the hyperpolarization (see Fig. (\ref{fig:Scheme})).

Polarization of spin environments has been largely dealt using resonant exchange interactions with highly polarized spins, the Hartmann-Hahn (HH) exchange method \cite{HHahn}. When this transfer happens from the highly polarized electron spin (due to its g-factor) to the nuclear spin, it is often dubbed dynamic nuclear polarization (DNP) \cite{DNPNV}. Heat-Bath algorithmic cooling(HBAC), provides an alternative
for cooling a target subset of spins \cite{schulman_scalable_1998, fernandez2004algorithmic}. 
These techniques 
use quantum operations to compress the entropy 
away from a target element in an   
ensemble of the spins to the rest of the ensemble and then 
through that to the heat-bath. 

For HBAC, it is assumed that there is an ensemble of spins or 
qubits and that universal quantum operations can be applied to them. 
The goal is to increase the polarization of a subset of spins in the 
ensemble, which are referred to as the ``computation'' qubits or spins. 
This idea was first introduced in \cite{schulman_scalable_1998} 
for closed systems and it was proposed to use 
entropy  compression algorithms to  push the entropy 
away from the target elements to the rest of spins which are 
known as the ``reset'' spins. 
However, the performance of the algorithm was limited to 
Shannon bound for compression. 
To circumvent the Shannon bound, Boykin et al. 
 proposed to use a heat-bath to transfer the compressed entropy out of
system \cite{boykin_algorithmic_2002}.   
In this setting, after the compression of the entropy, the reset spins, which are
now heated up,  
are cooled down in interaction with a heat-bath. 
Often this is the natural spin relaxation process that equilibrates the
reset spins, but it can also be actively done through optical 
pumping or similar techniques in some systems. 
After the reset spins are cooled down, the process of 
compression is repeated and this makes 
an iterative process where in each iteration, first the entropy 
is transferred away from the computation spins and compressed to 
the reset spins and then the reset spins are reset back to their 
initial state. 
However, even in the interaction with the heat-bath,  
HBAC techniques cannot always fully polarize the target spins. 
In  \cite{schulman_physical_2005}, Schulman et al.  showed that it is not always 
possible to completely purify the target spins and that the 
achievable cooling is limited. 
They introduced the Partner-Pairing 
algorithm (PPA), 
which, for compression, sorts the diagonal elements of the density matrix 
of the computation and reset spins decreasingly. They showed 
that this is the optimal HBAC technique and also found that 
it still cannot always reach full polarization 
for the target spins. 
Later, Raeisi and Mosca showed that PPA asymptotically converges and 
established the asymptotic state and found the cooling limit of all HBAC 
techniques \cite{raeisi2015asymptotic}.  

Here we report the first experimental realization of reaching the  
cooling limit of HBAC. We use a system of 
 three spins to 
polarize a Nitrogen nuclear spin in a nitrogen  vacancy centre in diamond. 
Since we can easily polarize the two carbon spins, 
we use both of them for the reset and the Nitrogen as the single computation spin. 
We use a technique similar to the one in \cite{ryan2008spin}, 
which consists of repeating a fixed set of operations 
in subsequent iterations. We show numerically that 
this method converges to the 
HBAC limit and verify this in our experiment.   

We start by describing our method of cooling and then present our 
results and compare them with PPA and the cooling limit of HBAC. 
We conclude with a brief discussion of the experiment  
and an outlook towards Quantum Thermodynamic applications.

We implement the circuit shown in Fig. 1 (b)with three nuclear spins viz., one  $14N$ spin  and two $13C$ spins. 
We choose the 14N as our traget nuclear spin to be hyperpolarized.Though the choice of the target qubit is not unique, a comparitively faster readout of the $14N$ allows us to reduce the overall experimental runtime \cite{qec}.
The carbon nuclear spins chosen for reset are strongly coupled with the electron spin, allowing  
for a faster and efficient swap of their polarization with the electron spin.

 The polarization of the two $13C$ spins  are reseted  in each iteration of the algorithm, which is depicted here by the $R$ box  and its experimental implementation in Fig.(\ref{fig:Scheme}-c). These reset operations are non-unitary gates  and are performed via the central NV spin, which we detail later.
Mathematically, we take the reset state of the the two carbon spins to be 
\begin{equation}
\rho_{C_{1,2}}=\frac{1}{z}\left(\begin{array}{cc}
e^{\eps_{1,2}} & 0\\
0 & e^{-\eps_{1,2}}
\end{array}\right).
\label{eq:reset-state}
\end{equation}
where $z$, is normalization constant. The spin-polarization's of the two carbon spins are then given by $P_{C_{1,2}} = \tanh{\epsilon_{1,2}}$.
The compression is done by the application of $\Uexp$ (shown in Fig.(\ref{fig:Scheme}-b)) which 
only swaps the $\mid 100\rangle$  and $\mid 011\rangle$ i.e., 
the target spin is flipped only when the states of the other two qubits 
are identical and dissimilar from the target spin state. When the  
circuit in Fig.(\ref{fig:Scheme}-b) is repeated, the target spin polarization 
grows 
and saturates to a value higher than the reset polarizations of the 
other two spins. We represent the final asymptotic state of  
the target spin with
\begin{equation}
\rho_{f}=\frac{1}{z}\left(\begin{array}{cc}
e^{\eps_{max}} & 0\\
0 & e^{-\eps_{max}}
\end{array}\right),\label{eq:Target-state}
\end{equation}
with $z$ the normalization factor. 
For the simple case of our three-qubit system, 
we can analytically calculate the asymptotic state of the $14N$ 
spin and show that it converges to the HBAC limit.
The limit of HBAC 
with reset polarizations $\eps_{1,2}$ is 
\begin{equation}
 \eps_{max} \leq 2^{n-1}(\eps_1+\eps_2), 
\end{equation} 
where $n$ is the number of computation qubits and 
 in our experiment, $n=1$. 
We can express the final polarization, $\pol_{max}$ in terms of $\eps_{max}$ as
\begin{equation}
\pol_{max} = p_0-p_1 = \frac{e^{\eps_{max}} - e^{-\eps_{max}}}{e^{\eps_{max}} + e^{-\eps_{max}}}
 = \tanh(\eps_{max}),
\end{equation}
where $p_0$ and $p_1$ are the probability of the spin  being in the up and down 
states correspondingly. 
This indicates that the polarizations  should be limited by \cite{raeisi2015asymptotic}
\begin{equation}
\pol_{max} = p_0-p_1 <  \tanh(\eps_1+\eps_2). 
\end{equation}
We also use $\pol_{C_1}$ and $\pol_{C_2}$ to show 
the polarization of the two Carbon nuclear spins. 
We experimentally obtain this for a single Nitrogen spin 
by repeating the above circuit 25 times, with 
the first qubit in the circuit being the 
Nitrogen and the second and the third being the carbon spins.  
We also include this limit in our plots for comparison.

\begin{figure}
\begin{centering}
\includegraphics[width=1.0\columnwidth,height=1.0\columnwidth]{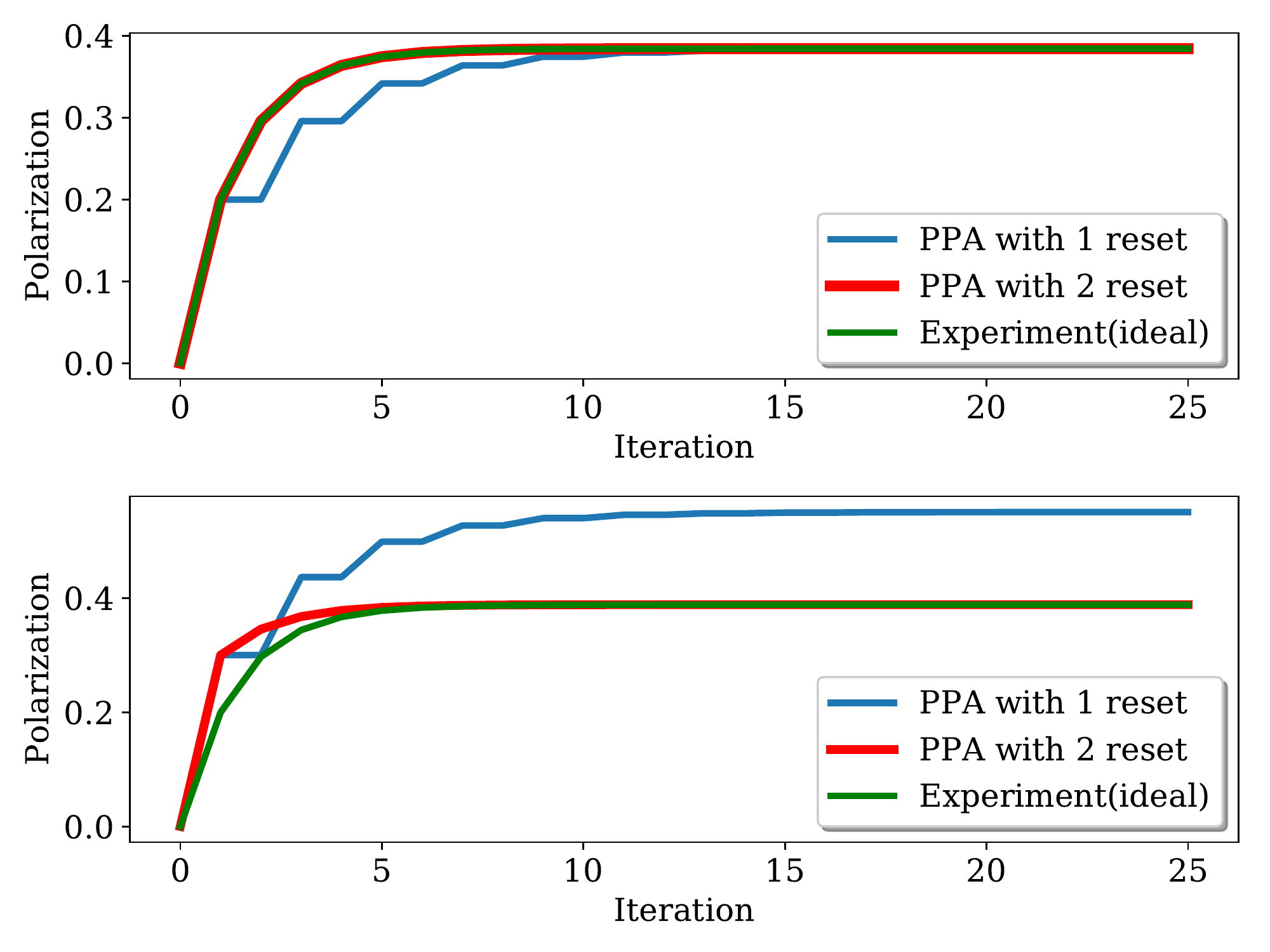} 
\par\end{centering}
\caption{\label{fig:Comparison}Theoretical comparison between
the PPA and our method (see Fig (\ref{fig:Scheme})). The 
plots show how the polarization of the target spin
would improve as more iterations are implemented. 
For the top plot, the  initial polarization of the two reset spins are $\pol_{C_1} = \pol_{C_2} =0.2 $ (top) and $\pol_{C_1}= 0.2, \pol_{C_2} = 0.3$ (bottom)
 }
\end{figure}

Figure (\ref{fig:Comparison}) compares the polarization build-up 
of the target Nitrogen spin in our method to PPA  
with single and two-qubit reset. 
For the single-qubit-reset  PPA, only one spin is reset and 
we assume that it is reset to $\eps_1>\eps_2$. 
For the top plot, we take the reset polarizations
 of both carbons  to be 
$\pol_{C_1,C_2} = 0.2$, whereas for the second plot, we set the reset 
polarization of the second qubit to $\pol_{C_2} =0.1$ and 
of the third qubit to $\pol_{C_1} =0.3$.
Note that PPA with single qubit reset can achieve 
higher polarization, compared to two-qubit-reset PPA. 
This is because with the two-qubit-reset, $\eps_2\leq \eps_1$  
and therefore the final polarization would be less. 

\begin{figure*}
\begin{centering}
\includegraphics[width=2\columnwidth,height=0.75\columnwidth]{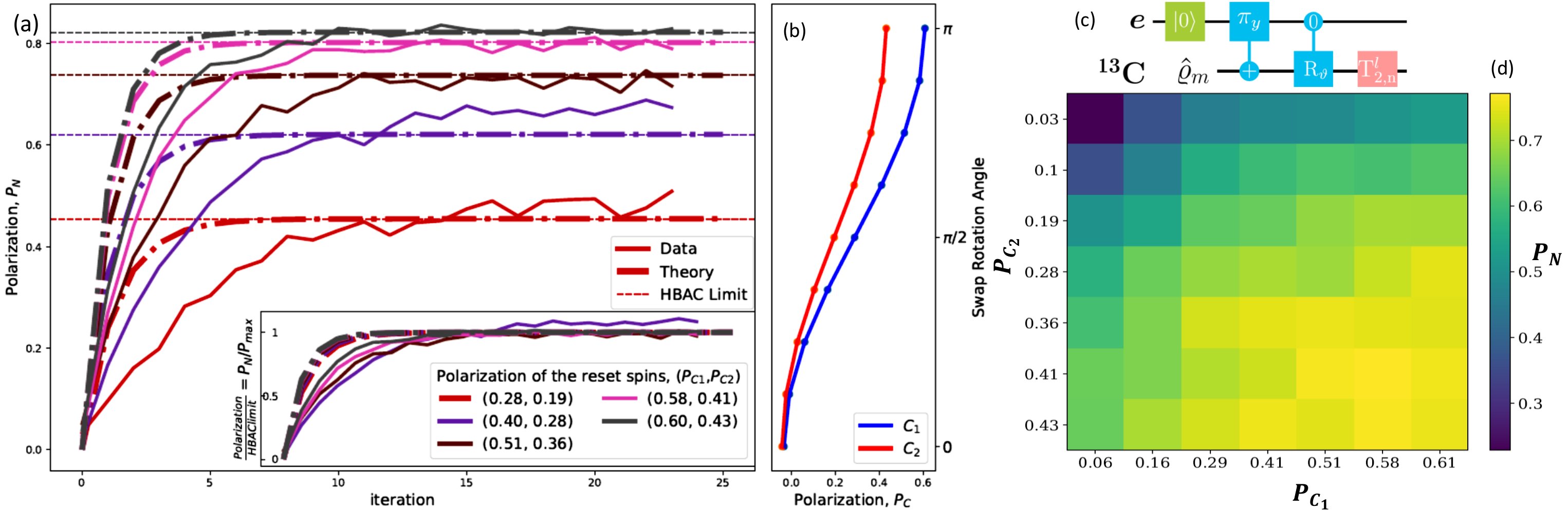} 
\par\end{centering}
\caption{\label{fig:Exp_result}(a) The hyperpolarization of 14N spins for varying reset polarizations on the two 13C nuclear spins is shown here. The experimental data is compared with the theoretical simulations and the ideal HBAC saturation limit. (Inset) In the inset we plot the ratio of experimental polarization of the $14N$, $P_N$ with the ideal HBAC limit $P_{max}$. (b) The initial polarization of two 13C  spins obtained  by varying the rotation angle $\theta$. (c) The quantum circuit employed for arbitrary initialization used in (a). (d) $3D$ plot showing the hyperpolarizaiton of the $14N$ over the polarization space of the two $13C$ spins.
 }
\end{figure*}

The experiments were carried out at room temperature on a homebuilt 
confocal microscope setup using a a type IIa CVD grown diamond 
crystal (layer) that has [100] surface orientation and a 13C 
concentration of $0.2\%$. The NVC is located $\sim 15 \mu$m 
below the diamond surface and a solid immersion lens has been 
carved around it via focused ion beam milling. A coplanar 
waveguide for microwave (MW) and RF excitation made from 
copper is fabricated onto the diamond via optical lithography. 
An external static magnetic field of $\sim 540$ mT from a 
permanent magnet is aligned along the symmetry axis of the 
NV centre (z-axis). A $532$ nm light is focused onto the NV 
by an oil immersion objective, which also collects light 
from the diamond. The fluorescence light of the NV is 
isolated by spatial and spectral filtering and finally 
detected with a single photon counting detector. Optical 
excitation of the NV centre polarizes the electron spin 
triplet (S=1) into its $m_S=0$ state, which forms the 
key element in the reset step shown in Fig.(\ref{fig:Scheme}), and the 
source of spin polarization in the experiment. 

Due to the weak coupling among the nuclear spins, no direct quantum gates among the nuclear spins, required for HBAC (Fig. 1(b)) can be performed. These gates are mediated by the electron spin, and hence is actively involved in the dynamics, (Fig. 1 (d)), in addition to being the polarization source for the carbon spins. The coupling of the NV electron to the surrounding nuclear spins is dominant along its symmetry (z) axis \cite{qec}.  The  $14N$ spin and the two nearest $13C$ spins of the lattice chosen for our experiments, are coupled the NV spin with strengths $-2.16$MHz, $90$kHz and $414$kHz respectively. 
While the two $13C$ spins are spin-$1/2$ systems the $14N$ is a spin-$1$ system with population distributed among the triplet levels $m_I = +1, 0, -1$ respectively. Hence, we choose a two-level subspace (TLS) formed by the $\ket{\pm 1}$ states of the $14N$ for our analysis, and omit the state $\ket{0}$, and renormalize the spin-state population in the TLS accordingly. The spin-polarization is then directly measured by the population difference $P_N = P_+-P_-$ within this effective TLS, as shown in Fig.3.

We transfer the optical spin polarization of the central (NV)-spin to the two $13C$ spins via SWAP gate which involves four conditional-NOT gates as shown in Fig.(\ref{fig:Scheme}-c). In the experiments, a nuclear spin reset (repolarization) is performed with a fidelity that is much higher than the total fidelity of the quantum circuit shown in Fig.(\ref{fig:Scheme}). This allows us to explore the whole polarization space ($\pol$) in HBAC. 

The three nuclear spins used in the experiments do not interact and thus each non-local gate performed among them needs to be mediated by the electron spin. Thus, unlike the reset operation, nuclear - nuclear swap gates occur via involvement of the electron spin and hence, suffer from electron $T_2$ decay, which is on the order of $500 \mu$s for this specific NV center \cite{bastipap}. This lowers the fidelity of the swap operation and the required time is at least as long compared to a reset of any nuclear spin.

After polarizing the two $13C$ spins using the SWAP gates shown in Fig.(\ref{fig:Scheme}-c), we perform the $U_e$ gate shown in Fig.(\ref{fig:Scheme}-d). This three-spin non-local gate swaps the population between the states $\ket{100} \leftrightarrow \ket{011}$, where the first spin state corresponds to the target $14N$ nuclear spin.
 We decompose $\Uexp$ to  three Toffoli gates that are further decomposed into $C_nNOT_e$ gates. Such gate have been earlier used in \cite{qec} for entangling the nuclear spins. 

We use optimal control platform DYNAMO \cite{dynamo} to realize fast electron spin ($2\pi$) rotations shown in Fig. 1 (d). As the individual hyperfine transitions are inhomogeneneously broadened to approximately 35kHz, it drastically complicates the implementation of electron spin gates required to have a spectral resolution better than 90kHz (the $A_{zz}$ coupling of the weakest coupled nuclear spin)\cite{bastipap}. Hence implementing electron spin rotations using shaped MW pulses, obtained from the optimization with DYNAMO, is key for our experiment \cite{qec}. The optimized pulses used here can avoid cross-talk between the individual electron hyperfine transitions even in a dense spectrum. The optimized  $U_e$ gate is then iterated until the $14N$ polarization $\pol_N$ saturates, which we show in Fig.(\ref{fig:Exp_result}-a). With the limitation from the decoherence time of the electron spin, and the decoherence induced by the electron optical excitation on the nuclear spins, the fidelity of the $U_e$ gates will eventually deteriorate with increasing number of iterations. In our experiment one iteration i.e., a reset operation and the non-local gate $\Uexp$ would take $\sim 570\mu$s. Given the various life times of the electron and nuclear spins \cite{qec}, $T_{2e} \approx 500\mu$s, and $T_{2n} \approx 8.5m$s, and the relaxation time of the nuclear spins $T_{2n} \sim 1$s we could safely perform $\sim 50$ iterations of the HBAC algorithm with high fidelity, which would allow the experimentally observed saturation of the polarization.

In the final part of our experiments, we use the ability to arbitrarily polarize the two reset spins to varying degree, and study the polarization gain of the $14N$ spin, as shown in Fig.(\ref{fig:Exp_result}-c). For this we first maximally polarize the two $13C$ spins via the SWAP gate with the electron spin, and then rotate the nuclear spins such that a superposition state, $ \sim \alpha \ket{0} + \beta \ket{1}$ is formed. By (laser-induced) dephasing of the carbon spin, one would obtain an ideal (thermal-like) polarization $P_C = |\alpha|^2 - |\beta|^2$. We would like to note that the HBAC, ideally is not going to be affected by the off-diagonal elements of the density matrix. Hence, imperfect dephasing of the nuclear spins (Fig. 3(c)) here, would not be a source of error in our experiments. Polarization is only affected by the diagonal elements of the density matrix and the operation $\Uexp$ would only permute the diagonal elements.  

In conclusion, we have experimentally 
implemented HBAC for three spins and reached the cooling limit of HBAC for this system. 
With 25 iterations, the Nitrogen spin reaches the $\pol_{max}$, 
the expected cooling limit of HBAC. 
The polarization of the $14N$ fluctuates  around the cooling limit 
and in some cases, it even exceeds the limit. 
This may be attributed to correlations between 
spins in the bath and the system \cite{rodriguez2017heat}, due to imperfections in $\Uexp$. 
But further experiments
are required to investigate this,and to benchmark the role of such correlations on the cooling limit of HBAC \cite{rodriguez2017heat}.

Further with the ability to polarize the reset spins to varying degree 
(a feature unique to our system), 
We benchmark our implementation of  HBAC for purification and polarization of the nuclear spins. 
The main requirement for using this method is that the spins in the system can be individually controlled 
and that the relaxation time of the target elements is long enough compared to the HBAC process. While the NV-based electron-nuclear spin system employed here is a good model system accommodating both these requirements, it can also be implemented in other solid-state systems to boost the polarization of a target spin system. Further, the achievable polarization would increase exponentially with employing more computation and reset spins. 

Combining our method with techniques like DNP would enhance the polarization of target spin(s) beyond that achievable directly with DNP. 

The physical system employed here represents a generic heat engine model for a quantum refrigerator, where the working fluid (electron spin) under modulation (quantum control) could continuously extract heat from a cold bath and dump it into the heat bath \cite{qrger}. 

\begin{acknowledgments}
        We would like to acknowledge the financial support by the ERC project SMeL, DFG (FOR2724), DFG SFB/TR21, EU ASTERIQS, QIA, Max Planck Society and the Volkswagenstiftung as well as the Baden-Wuerttemberg Foundation.
\end{acknowledgments}

\bibliographystyle{aipnum4-1}
\bibliography{HBAC}

\end{document}